\RequirePackage{fix-cm}
\documentclass[smallextended]{svjour3}       
\smartqed  
\usepackage{bm} 
\usepackage{url}  
\usepackage{graphicx}
\usepackage{color}
\newcommand{\cfujita}[1]{{#1}}

\begin{document}

\title{Numerical aspect of large-scale electronic state calculation for flexible device material
\thanks{The present research was partially supported by JST-CREST
project of 'Development of an Eigen-Supercomputing Engine 
using a Post-Petascale Hierarchical Model',  
Priority Issue 7 of the post-K project and KAKENHI funds (16KT0016,17H02828).
Oakforest-PACS was used through the JHPCN Project (jh170058-NAHI)
and through Interdisciplinary Computational Science Program in Center 
for Computational Sciences, University of Tsukuba.
The K computer was used in
the HPCI System Research Projects (hp180079, hp180219).
Several computations were carried out also on the facilities of the Supercomputer Center, the Institute for Solid State Physics, the University of Tokyo
and the Academic Center for Computing and Media Studies, Kyoto University.
}
}


\author{Takeo Hoshi \and
        Hiroto Imachi \and  Akiyoshi Kuwata \and Kohsuke Kakuda \and Takatoshi Fujita \and Hiroyuki Matsui
}


\institute{T. Hoshi A. Kuwata K. Kakuda \at
              Department of Applied Mathematics and Physics, Tottori University, 4-101 Koyama-Minami, Tottori, 680-8552, Japan  \\
              Tel.: +81-857-31-5448 \\
              Fax: +81-857-31-5747 \\
              \email{hoshi@damp.tottori-u.ac.jp}           
           \and
           H. Imachi \at
              Department of Applied Mathematics and Physics , Tottori University, 4-101 Koyama-Minami, Tottori, 680-8552, Japan \\
               \emph{Present address:  Preferred Networks, Inc.}   
           \and
           T. Fujita \at
              Department of Theoretical and Computational Molecular Science, Institute for Molecular Science, Japan \\
           \and
           H. Matsui \at
              Research Center of Organic Electronics, Yamagata University, Japan \\
 }

\date{Received: date / Accepted: date}

\maketitle

\begin{abstract}
Numerical aspects of large-scale electronic state calculation are explored 
on  flexible organic device materials. 
Physical theory, numerical method and real application studies 
are discussed in the context of application-algorithm-architecture co-design. 
An application study was carried out for disordered organic thin film. 
Participation ratio, a measure for the spatial extension of electronic wavefunction is focused on,
since it is crucial for device property.  
A data scientific research is reported for 
a classification problem of disordered organic polymers, 
in which participation ratio is used as descriptor.
These application studies indicate 
the potential need of purpose-specific solvers for internal eigenpairs.
\end{abstract}

\section{Introduction}

Large-scale quantum material simulation or electronic state calculation
is one of the major fields in computational science with supercomputers. 
A central problem in this field is 
the  generalized eigenvalue equation, in which 
an eigenvalue or eigenvector is 
the energy or the wavefunction of an electron, respectively. 
Although the {\it de fact} standard parallel solver library is 
ScaLAPACK \cite{SCALAPACK},  
these routines show severe limitations in parallel efficiency
and several scalable solvers, 
such as ELPA \cite{ELPA-URL}, \cite{ELPA2014} and EigenExa \cite{EigenExa-URL}, \cite{EigenExa-PAPER}, 
were developed recently.
Among them, 
ELPA was developed in the tight collaboration 
between numerical researchers and material researchers in Europe.  
\cite{FHI-AIMS,ELPA2014}
Such a fruitful collaboration requires
the co-design approach \cfujita{among} application, algorithm and architecture,
because an optimal algorithm is dependent both on problem and architecture.
For example, 
a ELPA paper \cite{ELPA2014} discusses
the benchmark of the calculation
that obtains all the eigenvalues and a small fraction (10-50 \%) of eigenvectors,
since such calculations are typical among electronic state calculations. 

The present paper reports 
large-scale electronic state calculations
in the context of application-algorithm-architecture co-design. 
The target application is organic flexible device materials,
and the application study contains (i) large single problem
and (ii) many small problems in data scientific research 
for the classification of disordered materials.
These researches lead us to a potential need for 
purpose-specific numerical solvers for internal eigenpairs. 
We believe that the present paper is a seed of  
the collaboration between numerical researcher and material researcher
with  the next-generation or (pre-)exascale supercomputers.

The present paper is organized as follows; 
Section ~\ref{SEC-ORGANIC} gives a brief overview of organic material.
Physical theory and related numerical method of large-scale electronic state calculation
appear in Sec.~\ref{SEC-BACKGROUND}. 
The calculated results appear for organic thin film in Sec.~\ref{SEC-APPLI-THINFILM} and
for organic polymer in Sec.~\ref{SEC-APPLI-POLYMER}\cfujita{.}
The potential need for purpose-specific solvers is discussed in Sec.~\ref{SEC-APPLI-NEED}\cfujita{.}
Section \ref{SEC-SUMMARY} is devoted to summary and future outlook.

\section{Organic materials \label{SEC-ORGANIC}}

Organic semiconductor material is the foundation of 
flexible devices, such as
flexible displays \cite{GELINCK-2004}, 
flexible solar cells \cite{XU-2018}, and human-friendly wearable
electronics \cite{SEKITANI-2011}. Unlike inorganic semiconductors such as silicon, an organic
semiconductor consists of small molecules or polymers which form solids with
weak van der Waals interaction.
An important industrial problem is 
to control the disorder of the atomic structure
without an increase of the fabrication cost. 
The electronic property is highly anisotropic and is governed by 
the spatial extension of the characteristic electrons called $\pi$ electrons that 
lie, for example, on benzene rings. 
A fundamental issue  is a conflicting demand on opt-electronic devices; 
An extended wavefuntion is preferable for conduction property, since an extended wavefunction can induce the electrical current easily. 
A localized wavefunction, by contrast, is preferable for optical property, 
since a localized wavefunction can strongly interact with light. The above conflicting demand should be a foundation of material design.
Wavefunctions in disordered structures are localized, while wavefunctions in ideal crystalline (periodic) structures 
are extended throughout the whole system.

An investigation on these materials requires
large-scale electronic state calculation in 10-100 nm scales, 
since complicated disordered structure is crucial for device properties.
The investigation requires also a data scientific research with capacity computation,
or simultaneous computation of many problems, 
because the real device property stem from
the average among many disordered samples. 

Here a crucial issue is to find a proper quantity that characterizes each disordered sample
in the context of device property. 
In the present paper,
the quantity is chosen to be 
participation ratio (PR)
\cite{PARTICIPATION-RATIO1}, \cite{PARTICIPATION-RATIO2}, 
\cite{PARTICIPATION-RATIO3}, \cite{PARTICIPATION-RATIO4},
\cite{PARTICIPATION-RATIO5}, 
the spatial extension of electronic wavefunctions,
since the spatial extension of wavefunctions governs the device property.
The detailed explanation of PR will be given in the next section.
Later in the present paper, 
PR will be used as descriptor  in a data scientific research for classification.

\section{Theory and numerical method 
\label{SEC-BACKGROUND}}
%

Numerical foundation of electronic state calculations is explained 
as the basics of the co-design approach. 
Details of the physical theory can be found in textbooks,
such as Ref.~\cite{MARTIN-TEXTBOOK}. 

\subsection{Physical origin of matrix problem}\label{sec:bp:phys:matrix}%

The fundamental Schr\"odinger-type equation, a partial differential equation in real space $\bm{r}$, 
is written for an electronic wavefunction $\phi(\bm{r})$ as 
\begin{eqnarray}
 \hat{H} \phi(\bm{r}) =\lambda \phi(\bm{r})
 \label{EQ-QM-EQN}
\end{eqnarray}
with  the Hamilton operator of 
\begin{eqnarray}
 \hat{H} \equiv - \frac{\hbar^2}{2m} \Delta + V_{\rm eff}(\bm{r}).
\end{eqnarray}
Here, 
$\Delta$ is Laplacian, 
$m$ is the mass of electron and $\hbar$ 
is the Planck constant, a physical constant ($\hbar \approx 1.05 \time 10^{-34}$Js). 
$V_{\rm eff}(\bm{r})$ is the effective potential, a scalar function. 
The normalization condition of 
\begin{eqnarray}
\int |\phi(\bm{r})|^2 d \bm{r} = 1  
 \label{EQ-NORMALIZAION}
\end{eqnarray}
is imposed and stems from the fact that the sum of the weight distribution of one electron 
should be the unity.  
The function of $n(\bm{r}) \equiv |\phi(\bm{r})|^2 (\ge 0)$ 
is the weight distribution of the electron at the point of $\bm{r}$.
The normalization condition of Eq.~(\ref{EQ-NORMALIZAION}) can be expressed as
\begin{eqnarray}
\int n(\bm{r}) d \bm{r} =1.
\label{EQ-NORMALIZAION-N}
\end{eqnarray}

An eigenvalue of $\lambda$ means the energy of an electron in the material
and is called eigenenergy. 
The $k$-th eigenpair of $(\lambda_k, \phi_k(\bm{r}))$ is defined for $k=1,2,..,M$
in the order of $\lambda_1 \le \lambda_2 \le \cdots \le \lambda_M$. 
Each material has a specific integer of $k_{\rm HO}$ called
highest occupied eigenenergy\cfujita{,} and the eigenpairs for $k=1,2,...k_{\rm HO}$ are occupied by the electrons. 
A para-spin material with $N_{\rm elec}$ electrons, for example, 
gives the value of $k_{\rm HO} = N_{\rm elec}/2$, if $N_{\rm elec}$ is even.
Semiconductor material has a finite energy gap between 
the $k_{\rm HO}$-th and $(k_{\rm HO}+1)$-th eigenenergies
($\lambda_{k_{\rm HO}+1} - \lambda_{k_{\rm HO}} >0$).

Now we consider, as a typical case, that  
$\phi(\bm{r})$ is expressed as a linear combination of given basis functions 
\begin{eqnarray}
 \phi(\bm{r}) = \sum_{j}^{M} v_j \chi_j(\bm{r}) 
 \label{EQ-QM-LCAO},
\end{eqnarray}
where $M$ is the number of the basis functions $\{ \chi_j(\bm{r}) \}$. 
The basis functions $\{ \chi_j(\bm{r}) \}$ are normalized to be
\begin{eqnarray}
\int \chi^\ast_j(\bm{r})  \chi_j(\bm{r}) d\bm{r} =1.
\end{eqnarray}
A typical function is called atomic orbital
and is localized near the position of an atomic nucleus. 
Since each basis function belongs to one atom,
the basis index $i$ is equivalent to the composite indices
of \cfujita{an atom index $I$ and an orbital index $\alpha$ ($i \equiv (I, \alpha)$).
The orbital index $\alpha$ distinguishes the basis functions that belong to the same atom but different in their shape. }

A generalized eigenvalue equation appears,
when  Eq~(\ref{EQ-QM-LCAO}) is  substituted for
Eq.~(\ref{EQ-QM-EQN}); 
\begin{eqnarray}
 A \bm{v} = \lambda B \bm{v}
 \label{EQ-QM-GEP}
\end{eqnarray}
with the $M \times M$ matrices of
\begin{eqnarray}
 A_{ij} &\equiv& \int \chi_i^\ast(\bm{r}) \hat{H}  \chi_j(\bm{r}) d\bm{r} \\
 B_{ij} &\equiv& \int \chi_i^\ast(\bm{r})  \chi_j(\bm{r}) d\bm{r}. 
\end{eqnarray}
The matrices $A$ and $B$ are Hermitian.
The matrix $B$ is positive definite and 
satisfies $B_{jj}=1$ and $| B_{ij}| < 1$($i \ne j$).
Hereafter we consider, as among many researches, 
that the basis functions are real and  
the matrices $A$ and $B$ are real-symmetric. 
The normalization condition of Eq.~(\ref{EQ-NORMALIZAION}) is reduced to 
\begin{eqnarray}
\bm{v}^{\rm T} B  \bm{v} = 1,
\label{EQ-B-NORMALIZATION-V}
\end{eqnarray}
which is called $B$-normalization.
The present paper will discuss matrix data generated 
by our simulation software ELSES \cite{ELSES-URL}, \cite{ELSES-2012},
an electronic-state calculation software
with first-principles-based modeled
(tight-binding) electronic-state theory. 
Sparsity of the matrices of $A_{ij}$ and $B_{ij}$ are explained briefly.
As explained in the previous subsection, the indices $i$ and $j$ are the composite indices of the atom indices $I$ 
and $J$ and the orbital indices $\alpha$ and $\beta$, respectively ($i \equiv i(I,\alpha), j \equiv j(J,\beta))$).
Therefore, an element of the matrices $A$ and $B$ is expressed by the four indices as $A_{I\alpha;J\beta}$ and $B_{I\alpha;J\beta}$, respectively.
Since a matrix element value decreases quickly and monotonically as the function of the inter-atomic distance between the $I$-th and $J$-th atoms ($r_{IJ}$), a cutoff distance $r_{\rm cut}$ \cfujita{can be} introduced.
A matrix element, $A_{I\alpha;J\beta}$ or $B_{I\alpha;J\beta}$, is ignored, if $r_{IJ} > r_{\rm cut}$, which makes the matrices to be sparse.

\subsection{Mulliken weight \label{SEC-MULLIKEN-WEIGHT}}

This subsection introduces 
Mulliken weight \cite{MULLIKEN-CHARGE}, 
a famous discretized representation for 
the weight distribution of $n(\bm{r})$. 
When the quantity of $q_i^{\rm(bas)}$ is defined as
\begin{eqnarray}
q_i^{\rm(bas)} \equiv \sum_j v_i B_{ij}  v_j,
\label{EQ-MULLIKEN-Q-I}
\end{eqnarray}
for $i=1,2,....,M$,   
it is called Mulliken weight at the $i$-th basis function.
The normalization condition of 
Eq.~(\ref{EQ-B-NORMALIZATION-V}) is reduced to
\begin{eqnarray}
\sum_i q_i^{\rm(bas)} = 1
\label{EQ-MULLIKEN-Q-I-SUM},
\end{eqnarray}
which is analogous to Eq.~(\ref{EQ-NORMALIZAION-N}). 

Since the basis index of $i$ is the composite indices of the atom index of $I$ and 
the orbital index of $\alpha$ ($i \equiv (I, \alpha)$), 
the Mulliken weight at the $I$-th atom is defined as
\begin{eqnarray}
q^{\rm(atm)}_I \equiv \sum_\alpha q_{I, \alpha}^{\rm(bas)}
\label{EQ-MULLIKEN-ATOM-I}
\end{eqnarray}
and satisfies 
\begin{eqnarray}
\sum_I^{\rm (atm)}
q^{\rm(atm)}_I =1\cfujita{.}
\end{eqnarray}
If each atom belongs to one molecule,
the Mulliken weight at the $P$-th molecule is defined as
\begin{eqnarray}
q^{\rm(mol)}_P \equiv \sum_I^{{\rm (atm) } \in P} q^{\rm(atm)}_I,
\label{EQ-MULLIKEN-Q-I}
\end{eqnarray}
where $\sum_I^{{\rm (atm) } \in P}$ is the summation among the atoms
that belong to the $P$-th molecule. The sum is the unity; 
\begin{eqnarray}
\sum_P^{\rm (mol)}
q^{\rm(mol)}_P=1. 
\end{eqnarray}
In this paper, we will use
the definition of 
$\bm{q}^{\rm(bas)} \equiv (q_1^{\rm(bas)}, q_2^{\rm(bas)}, ...,q_M^{\rm(bas)})^{\rm T}$
and
$\bm{q}^{\rm(mol)} \equiv (q_1^{\rm(mol)}, q_2^{\rm(mol)}, ...,q_\mu^{\rm(mol)})^{\rm T}$,
where $\mu$ is the number of the molecules in the system. 

\subsection{Participation ratio \label{SEC-PR-INTRO}}

This subsection introduces 
participation ratio (PR),
as a measure of the spatial extension of wavefunctions $\phi(\bm{r})$
or how the wavefunction spreads in real space
\cite{PARTICIPATION-RATIO1}, \cite{PARTICIPATION-RATIO2}, 
\cite{PARTICIPATION-RATIO3}, \cite{PARTICIPATION-RATIO4}.
Since the spatial extension of wavefunction governs the electrical conductivity,
PR was used with large-scale electronic state calculations, such as 
a research on the anomalous electrical conductivity in 
quasi crystals \cite{PARTICIPATION-RATIO5}.

In the continuum representation, 
PR is defined for a wavefunction $\phi(\bm{r})$ as
\begin{eqnarray}
  P^{\rm(cnt)(4)}(\phi) \equiv \left( \int |\phi(\bm{r})|^4 d \bm{r} \right)^{-1}.
\end{eqnarray}
under the $L^2$-normalization of Eq.~(\ref{EQ-NORMALIZAION}). 
For example, 
suppose $D$ is a closed area whose volume is $\Omega$ and $\phi$ is constant in $D$ as
\begin{eqnarray}
\phi(\bm{r}) =
\left\{
\begin{array}{ll}
\frac{1}{\sqrt{\Omega}} & (\bm{r} \in D)\\
0 & (\rm{otherwise}). 
\end{array}
\right.
\end{eqnarray}
Then the PR of $\phi$ gives the volume of the non-zero region of $\phi$ 
\begin{eqnarray}
   P^{\rm(cnt)(4)}(\phi) = \left( \frac{1}{\Omega^2} \int_D d \bm{r} \right)^{-1} = \Omega.
\end{eqnarray}
The above definition of PR can be expressed also 
by the weight distribution of $n(\bm{r})$ as
\begin{eqnarray}
 P^{\rm(cnt)(2)}(n) \equiv \left( \int |n(\bm{r})|^2 d \bm{r} \right)^{-1}
\end{eqnarray}
under the normalization of Eq.~(\ref{EQ-NORMALIZAION-N}).

PR for the eigenvector of $\bm{v}\equiv(v_1, v_2,....,v_M)^{\rm T}$ 
in Eq.~(\ref{EQ-QM-GEP}) can be also defined 
in a similar manner.   
The definition in the present paper is 
\begin{eqnarray}
P^{(4)}(\bm{v}) \equiv \left( \sum_j{|v_j|^4} \right)^{-1}
\label{EQ-DEF-PR-ORG}
\end{eqnarray}
under the $B$-normalization constraint of Eq.~(\ref{EQ-B-NORMALIZATION-V}). 
For the discretized representation,
PR indicates a measure of
the number of non-zero elements, namely, how broadly the elements exist on the indices.
For example,
the case of $B=I$ and
\begin{eqnarray}
\bm{v} \equiv \left(\frac{1}{\sqrt{3}}, \frac{1}{\sqrt{3}}, \frac{1}{\sqrt{3}}, 0, 0, ..., 0 \right)^{T},
\end{eqnarray}
the PR is 
\begin{eqnarray}
P^{(4)} = \left\{ \sum_j |v_j|^4 \right\}^{-1}
= \left\{ 3 \left( \frac{1}{\sqrt{3}} \right)^4 \right\}^{-1}
= 3.
\end{eqnarray}
Another definition is the one for the Mulliken weight on basis function as
\begin{eqnarray}
P^{(2)}(\bm{q}^{\rm(bas)}) \equiv \left( \sum_j{|q_j^{\rm(bas)}|^2} \right)^{-1}
\label{EQ-DEF-PR-MULLIKEN}
\end{eqnarray}
or the one for the Mulliken weight on molecules as 
\begin{eqnarray}
P^{(2)}(\bm{q}^{\rm(mol)}) \equiv \left( \sum_P^{\rm (mol)}{|q_P^{\rm(mol)}|^2} \right)^{-1}
\label{EQ-DEF-PR-MULLIKEN-MOL}. 
\end{eqnarray}
The quantity of Eq.(\ref{EQ-DEF-PR-MULLIKEN-MOL}) 
is called \lq molecular PR' in this paper and 
will appear later in this paper.

\subsection{Sparsity of matrix data}

According to the physical origin, 
the sparsity in matrix data of $A$ and $B$ is determined
by the atomic structure of the simulated system.
If a system contains independent (non-interacting) molecules,
for example, the matrices of $A$ and $B$ are block diagonal 
and\cfujita{,} each block stems from a molecule. 
\cfujita{The} sparsity of the matrices is crucial for
the efficiency of sparse-matrix solvers\cfujita{.}
\cfujita{A} matrix data library of \lq ELSES matrix library' \cite{ELSES-MATRIX-LIB} 
was constructed,
so as to enhance the collaboration between material and numerical researchers. 
The sparsity is different among the matrix data,
as seen in Fig.~3 of Ref.~\cite{K-EP}, for example. 
Two examples are explained;
The matrix data of \lq APF4686' stems 
from an organic polymer system, poly-(9,9 dioctyl-fluorene),
in a disordered structure
with 2076 atoms \cite{ELSES-2012}, \cite{ELSES-2014}.
The matrix size is $M=4686$ and
the number of non-zero elements is $N_{\rm NZ}=53950$.
The ratio of non-zero elements is $\gamma \equiv N_{\rm NZ}/M^2 \approx 0.0025$.
On the other hand, the matrix data of \lq AUNW9180' stems
from a disordered multishell gold nanowires with 1020 atoms
\cite{HOSHI-2009-AUNW}.
The matrix size of is $M=9180$ and
the number of non-zero elements is $N_{\rm NZ}=1783313$.
The ratio of non-zero elements is $\gamma \equiv N_{\rm NZ}/M^2 \approx 0.021$. 
It is noteworthy that ELSES matrix library has several features 
for the convenience of numerical researchers;
(i)  The matrix data files are recorded in the Matrix-Market format.
\cite{MATRIX-MARKET}
(ii) Eigenvalues and PR values are stored as files,
as well as the matrix data.

\begin{figure*}
  \includegraphics[width=0.95\textwidth]{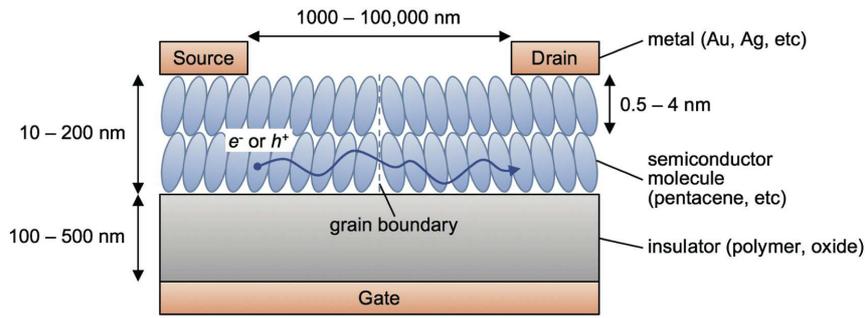}
\caption{Schematic structure and typical scales of organic field-effect transistors.  
The electrical current is depicted as an arrow squiggle. 
}
\label{FIG_DEVICE}       
\end{figure*}

\section{Result \label{SEC-APPLI} }

The present section is devoted to the results of 
large-scale electronic state calculations of organic semiconductor materials. 
The present paper focuses on p-type semiconductor, in which 
the electrical current stems from a small number of wavefunctions
that has the internal eigenenergies $\lambda_k$ near the highest occupied one $\lambda_{\rm HO}$
($\lambda_k \le \lambda_{\rm HO}$). 
Therefore, one should calculate only these internal eigenpairs.  

We calculated 
thin film or two-dimensional condensed organic semiconductor molecules in disordered structures, 
which is a proto-typical system of organic field-effect transistors (OFETs) shown schematically in Fig.~\ref{FIG_DEVICE}. 
OFETs consist of four layers: a gate electrode layer, an insulating layer, an organic semiconductor layer, 
and a source/drain electrode layer, typical scales of which are shown in Fig.~\ref{FIG_DEVICE}. 
The details of  Fig.~\ref{FIG_DEVICE} is explained in textbooks, like Ref.~\cite{OFET-TEXTBOOK}. 
In the OFETs, the electrical current flows on several atomic semiconductor layers on the interface region between the polycrystalline 
semiconductor and amorphous insulator layers. So we should investigate a thin film system with disordered structures.

The present calculated system is a thin film of pentacene molecules (C$_{22}$H$_{14}$). 
Pentacene is one of the most famous organic semiconductors.
The present study is motivated by an experimental data of electron spin resonance (ESR) experiment for 
pentacene OFETs \cite{MATSUI-ESR-2010,MATSUI-ESR-2013}. 
The analysis of hyperfine interaction between hole carriers and protons in pentacene thin film 
in ESR spectra gives a molecular PR defined in Eq.(\ref{EQ-DEF-PR-MULLIKEN-MOL}) \cite{MATSUI-ESR-2010}. 
The molecular PR is denoted as $P$ hereafter. 
The experimental data indicates 
the appearance of \lq semi-localized' wavefunction that is extended among a few tens of molecules 
$(P = O(10))$. 
Such semi-locality is crucial for the device performance.

The generalized eigenvalue problem was solved by the mini-application of EigenKernel 
\cite{IMACHI-JIT2016}, \cite{EIGENKERNEL-URL}, \cite{EIGENKERNEL-2018}. 
EigenKernel is a middleware for the various solvers 
in ScaLAPACK, ELPA, and EigenExa and their hybrids. 
Although we have developed a massively parallel electronic state calculation method without eigenvalue problem
\cite{ELSES-URL}, \cite{ELSES-2012},  \cite{HOSHI2016SC16}, 
we still need eigenvalue solver, so as to obtain eigenpairs in the discussion of electronic property.

\subsection{Large-scale calculation of thin film organic material \label{SEC-APPLI-THINFILM}}

The calculated system is a thin-film (single molecular layer) system
with an artificial two-dimensional periodic simulation cell. 
The simulation cell contains $N_{\rm mol}$=1800 molecules\cfujita{,}
and the matrix size of the generalized eigenvalue problem is $M=183600$. 
Thousands of molecules are required in the simulation cell, 
so as to observe a semi-localized state extended among tens of molecules 
($N_{\rm mol} \gg O(10)$).

\begin{figure*}
  \includegraphics[width=0.95\textwidth]{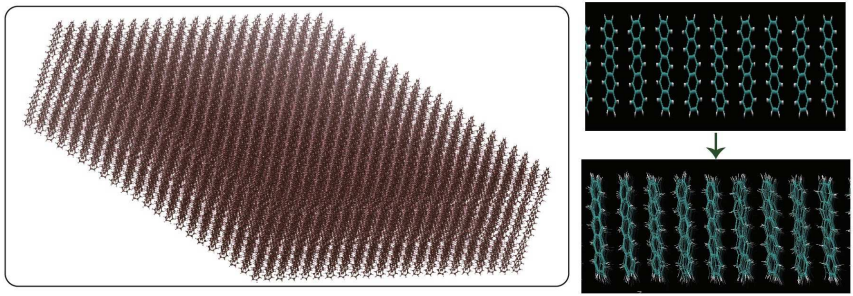}
\caption{Preparation of disordered pentacene thin film sample. (a) The initial structure 
of a three-layered sample in the crystalline geometry.
(b) The result of finite-temperature simulation.  
}
\label{FIG_PENTF_MD}       
\end{figure*}

The initial atomic structure of the disordered thin film sample 
was generated in classical molecular
dynamics simulations by GROMACS 
\cite{GROMACS-URL}, \cite{GROMACS}
with the generalized AMBER force field (GAFF) \cfujita{parameter set} \cite{GAFF}. 
The software and parameter set are standard for organic materials. 
The sample was prepared in the following stages; 
(I) A layered structure of pentacene in the crystalline geometry 
was prepared with three layers, shown in Fig.~\ref{FIG_PENTF_MD}(a). 
(II) Several finite temperature simulations were carried out 
at the temperature of 300-1000 K for the dynamics in 0.1-100 ns,
so as to generate a finite-temperature disordered structure,
as shown in Fig.~\ref{FIG_PENTF_MD}(b). 
In the simulation,
only the middle layer in Fig.~\ref{FIG_PENTF_MD}(a) was set to be mobile, and 
the upper and lower layers were set to be fixed,
so as to impose the boundary condition on the middle layer. 
The temperature of the simulation should be distinguished from the experimental one,
owing to the boundary condition and 
we have not yet compared the detailed results among different temperature and/or simulation time. 
A long-time (nano-second) dynamics is difficult 
for quantum simulation, owing to huge CPU time 
and we used the classical simulation.
Classical simulations, however, do not treat electronic wavefunction,
and the electronic state calculation is required for the disordered structure, 
so as to obtain wavefunctions. 
The total computational time of
the classical molecular dynamics simulation 
is $T_{\rm comp} \approx $ 5 hours for a nano-second dynamics by six nodes of a Intel-Xeon-based Supercomputer
at the Academic Center for Computing and Media Studies, Kyoto University.
The time of the quantum simulation is $T_{\rm comp} \approx $ 6 hours by 36 nodes
of a Intel-Xeon-based Supercomputer
at the  Institute for Solid State Physics, the University of Tokyo,
when the ScaLAPACK solver was used.  
It is noteworthy that 
the computational time of the quantum simulation with a dense-matrix solver
is proportional to $N_{\rm mol}^3$ and will be severe for a larger sample,
while the time of the classical simulation is proportional to $N_{\rm mol}^d$ with $1 \le d \le 2$.

\begin{figure*}
  \includegraphics[width=0.95\textwidth]{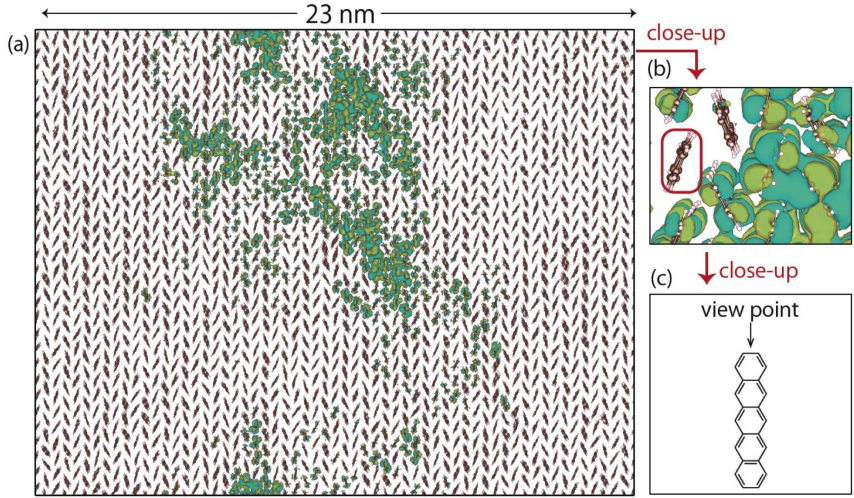}
\caption{Example of semi-localized wavefunction that appears on a pentacene thin-film system. 
The wavefunction $\phi(\bm{r})$ has the eigenenergy of $\lambda \approx - 0.15$eV,
where the energy origin $(\lambda \equiv 0)$ is set to be the eigenenergy 
of the highest occupied wavefunction. 
The wavefunction of $\phi(\bm{r})$ is depicted as two iso-surfaces in the opposite signs 
($\phi(\bm{r})=\pm C$, where $C$ is a positive constant).  
The two surfaces are painted by different colors. 
(a) The whole region of the periodic simulation cell. (b) A close-up of (a). 
(c) A picture of a single pentacene molecule with the view point of (b).  
}
\label{FIG_PENTF_WFN}       
\end{figure*}

Figure \ref{FIG_PENTF_WFN} shows a typical wavefunction with the molecular PR of $P \approx 35$,
which agrees with the experimental observation of semi-localized wavefunction ($P = O(10)$). 
We should say, however, that the present simulation 
is the one for an isolated thin film system of pentacene layer 
and ignores the affects the neighboring layers,
whereas the experimental result \cite{MATSUI-ESR-2010,MATSUI-ESR-2013} indicates
the importance of the neighboring insulator layer shown in Fig.~\ref{FIG_DEVICE}.
Now the calculations are on going
for  the interface system including pentacene layer and insulator layers.

\subsection{Data scientific research of organic polymers \label{SEC-APPLI-POLYMER}}

This subsection gives a data scientific research with capacity computation,
or simultaneous computation of many middle-size problems.
The purpose of the research is how to characterize 
the disordered structure in the context of device property.
In general, the spatial extension of wavefunction 
plays a crucial role on device property and is rigorously measured by
PR for each wavefunction. 
Therefore a set of PR values among wavefunctions in a sample
can be a candidate of the measure of the disorder for the sample.
In other words, the set of PR values can be a descriptor 
that characterizes the sample. 

The present paper focuses on a research on
poly-(phenylene-ethynylene) (PPE) ~\cite{TERAO-2013},
a typical conducting polymer.
The paper measures the device property or the mobility
for isolated polymers and found the importance of structural disorder.  
The first stage of theoretical research on disordered polymers 
is to define the descriptor of disordered polymers.
The present paper proposes that 
the set of PR values can give a descriptor for a disordered sample. 

\begin{figure}[htb]
  \centering
   \includegraphics[width=0.5\textwidth]{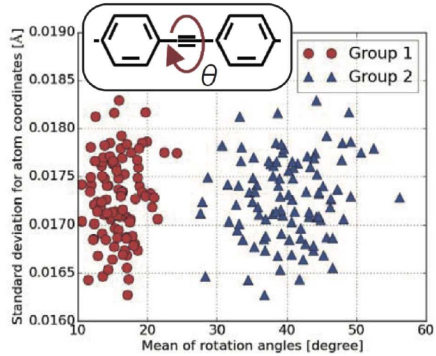}
   \caption{K-means clustering of polymer structures by their eigenvector participation ratios.
     Each structure is plotted by two statistical quantities.
     The x-axis is the mean of the rotation angles between adjacent benzene rings.
     The y-axis is the resulting standard deviation of the small fluctuation to the coordinates.
     Two markers, red circle and blue triangle, correspond to the two clusters labeled by the K-means algorithm.
  }
  \label{fig-kmeans}
\end{figure}

Figure \ref{fig-kmeans} shows a classification problem  among $N_{\rm sample}=200$ disordered PPE polymer samples
by the K-means clustering method, a typical classification algorithm. 
The structure of PPE polymers consists of 240 atoms\cfujita{,} and a part is drawn in the inset of Fig.~\ref{fig-kmeans}.
A polymer sample consists of 20 benzene rings.
The structural disorder \cfujita{was} introduced in 
the relative rotation angles between adjacent benzene rings shown in $\theta$ in Fig.~\ref{fig-kmeans}. 
The angles \cfujita{were} set from the normal distribution with standard deviation of 
20 or 60 degree\cfujita{. The samples were} generated 100 times for each class (total $N_{\rm sample} = 200$ samples).
After the rotations, small fluctuations taken from the normal distribution with standard deviation of 0.01 
\AA \cfujita{~were} given to all the coordinates of all atoms to remove degeneracy.
For each  sample $(i = 1, \dots, N_{\rm sample})$, 
the generalized eigenvalue problem  of
\begin{eqnarray}
  H^{(i)} \bm{v}^{(i)}_{j} = \lambda^{(i)}_{j} S^{(i)} \bm{v}^{(i)}_{j} \ \ (j = 1, \dots, M).
\end{eqnarray}
\cfujita{was} solved numerically with the matrix size of $M = 714$.
Here, the list of PR value of all the eigenvectors 
\begin{eqnarray}
  \bm{d}^{(i)} := (P^{(4)}(\bm{v}^{(i)}_{1}), \dots, P^{(4)}(\bm{v}^{(i)}_{M})))^T
\end{eqnarray}
\cfujita{was} used as the descriptor vector for the $i$-th sample.

The classification in 
the K-means clustering algorithm was carried out
with the descriptor vectors of $\{ \bm{d}^{(i)} \}_{i=1, \dots, M}$. 
The number of clusters was two.
The classification result is shown in Fig.~\ref{fig-kmeans}.
Each sample is plotted on the plane by two statistical quantities
with markers corresponding to the cluster labels given by the K-means algorithm.
As a result, the two clusters perfectly matched to the structure classes.
Namely, all the samples with adjacent rotation angles in standard deviation of 20 degree are clustered into one group,
and all the samples with adjacent rotation angles in standard deviation of 60 degree are clustered into the other group.
We should note the generality of PR as descriptor,
since PR is calculated uniquely for any material without any preknowledge. 
The present result implies that
the PR of wavefunctions can be a important quantity that bridges 
between disordered structure and device property. 
A more extensive study with principal component analysis is on going \cite{PCA-REPORT}
and will be reported elsewhere. 
It is noted that the present data scientific research is one among single polymer samples and
a challenging future problem is the data scientific research for condensed polymer samples,
as a foundation of device material research.

\subsection{Potential need of purpose-specific solvers \label{SEC-APPLI-NEED}}

The application studies in the previous sections 
are carried out by the dense-matrix solver and
implies the potential need of novel solvers. 
The largest matrix size in the present paper is $M=2 \times 10^5$.
In general, 
the computational cost of dense-matrix solvers is proportional to $M^3$ and
a previous paper \cite{HOSHI2016SC16} reported
that the dense-matrix solver solved
a million dimensional generalized eigenvalue problem ($M=10^6$) 
by 1.5 hours with the whole system of the K computer. 
Since the above calculation indicates the practical upper limit of the dense-matrix solver,
novel solvers are needed, at least,  for a problem with $M \ge 10^6$. 

Here we discusses 
the potential need of purpose-specific solvers suitable to the present problem.
The need is the one for the solver of internal eigenpairs, 
like z-PARES \cite{Z-PARES},\cite{Z-PARES-URL}, FEAST \cite{FEAST},\cite{FEAST-URL}, 
the filtering method \cite{FILTERING}, k-ep \cite{K-EP}, \cite{K-EP-URL},
because we would like to calculate only internal eigenpairs
with the eigenenergies $\lambda_k$ near the highest occupied one $\lambda_{\rm HO}$
($\lambda_k \le \lambda_{\rm HO}$). 
Internal eigenpair solvers are desirable both for a solution of large problems,
or a faster solution of middle-size problems. 
Application researchers, however, find sometimes the difficulty in choosing a solver,
since the performance can be dependent on problem and machine.
For example, the performance is dependent not only the matrix size $M$ but also the sparsity,
when one uses a sparse-matrix solver. 
A possible remedy for the difficulty is to develop a \lq middleware' that 
provides a universal interface for various solver routines,
like EigenKernel.

\section{Summary and future outlook \label{SEC-SUMMARY}}

The present paper discusses
the generalized eigenvalue problem 
in large-scale electronic state calculation for  flexible organic device materials. 
The application studies were carried out for disordered organic thin film and  polymer. 
The calculation of participation ratio is focused on, 
since it is a measure of the spatial extension of electronic wavefunctions
and governs the device property. 
The present application research indicates the potential need of purpose-specific solvers 
with internal eigenpairs. 

\section*{Acknowledgement}
The authors thank to  Tomofumi Tada (Tokyo institute of Technology)
and Jun Terao (University of Tokyo) 
for fruitful discussion on organic polymer.



\begin{thebibliography}{}
%
%

\bibitem{SCALAPACK}
ScaLAPACK: \url{http://www.netlib.org/scalapack/}

\bibitem{ELPA-URL}
ELPA: \url{http://elpa.mpcdf.mpg.de/}

\bibitem{ELPA2014}
Marek, A., Blum, V., Johanni, R., Havu, V., Lang, B., Auckenthaler, T., Heinecke, A., Bungartz, H.~J.~and Lederer, H., 
\lq The ELPA Library -- Scalable Parallel Eigenvalue Solutions for Electronic Structure Theory and Computational Science',
J.~Phys.~Condens.~Matter {\bf 26} (2014) 213201.

\bibitem{EigenExa-URL}
EigenExa: \url{http://www.r-ccs.riken.jp/labs/lpnctrt/en/projects/eigenexa/}

\bibitem{EigenExa-PAPER}
Imamura, T., Hirota, Y.,  Fukaya, T., Yamada, S., and Machida, M.
\lq EigenExa: high performance dense eigensolver, present and future', 8th
International Workshop on Parallel Matrix Algorithms and Applications
(PMAA14), Lugano, Switzerland, 2014.

\bibitem{FHI-AIMS}
Blum, V., Gehrke, R., Hanke, F., Havu, P., Havu, V., Ren, X., Reuter, K. and Scheffler, M.:
Ab initio molecular simulations with numeric atom-centered orbitals, 
Computer Physics Communications {\bf 180} (2009) 2175-2196;
\url{https://aimsclub.fhi-berlin.mpg.de/}

\bibitem{GELINCK-2004}
Gelinck,  G. H.,  
Huitema, H. E., 
van Veenendaal, E., 
Cantatore, E., 
Schrijnemakers, L.,
van der Putten, J. B.,
Geuns,  T. C.,
Beenhakkers,  M.,
Giesbers,  J. B.,
Huisman,  B. H.,
Meijer,  E. J.,
Benito,  E. M., 
Touwslager,  F. J.,
Marsman,  A. W.,
van Rens,  B. J.,
de Leeuw,  D. M., 
\lq Flexible active-matrix displays and shift registers based on solution-processed organic transistors',
Nat. Mater. 3 (2004) 106-110.

\bibitem{XU-2018}
Xu, X., Fukuda, K., Karki, A.,  Park, S., Kimura, H., Jinno, H., Watanabe, N., 
Yamamoto, S., Shimomura, S., Kitazawa, D., Yokota, T., Umezu, S., 
Nguyen, T-Q., and Someya, T., 
\lq Thermally stable, highly efficient, ultraflexible organic photovoltaics', 
PNAS 115 (2018) 4589-4594.

\bibitem{SEKITANI-2011}
Sekitani, T.,  and Someya, T.,  
\lq Human-friendly organic integrated circuits', 
Mater. Today 14 (2011) 398-407.

\bibitem{PARTICIPATION-RATIO1}
Bell, R. J.,  and Dean, P., 
\lq Atomic vibrations in vitreous silica',
Disc. Faraday Soc. 50, (1970) 55-61.

\bibitem{PARTICIPATION-RATIO2}
  Bell, R. J., 
  \lq The dynamics of disordered lattices',
  Rep. Prog. Phys. 35 (1972)  1315.

\bibitem{PARTICIPATION-RATIO3}
  Thouless, D. J., 
  \lq Electron in disordered systems and the theory of localization',
  Phys. Rep. 13 (1974) 93-142.

\bibitem{PARTICIPATION-RATIO4}
  Wegner, F.,
  \lq Inverse Participation Ratio in 2 + $\varepsilon$ Dimensions',
  Z. Physik B 36 (1980) 209-214.

\bibitem{PARTICIPATION-RATIO5}
 Fujiwara, T., Mitsui, T., and Yamamoto, S., 
 \lq Scaling properties of wave functions and transport coefficients in quasicrystals',
 Phys. Rev. B 53 (1996) R2910-R2913.


\bibitem{MARTIN-TEXTBOOK}
Martin, R. M., 
\lq Electronic Structure - Basic Theory and Practical Methods', 
Cambridge University Press (2004). 

\bibitem{ELSES-URL}
ELSES: \url{http://www.elses.jp}

\bibitem{ELSES-2012}
  Hoshi, T., Yamamoto, S., Fujiwara, T., Sogabe, T.~and Zhang, S.-L.:
  An order-$N$ electronic structure theory with generalized eigenvalue equations 
  and its application to a ten-million-atom system,  
  J.~Phys.~Condens.~Matter {\bf 24} (2012) 165502.

\bibitem{MULLIKEN-CHARGE}
Mulliken, R. S. 
\lq Electronic Population Analysis on LCAO-MO Molecular Wave Functions. I',
J. Chem. Phys. 23, (1955) 1833-1840. 


\bibitem{ELSES-MATRIX-LIB}
ELSES matrix library:  
\url{http://www.elses.jp/matrix/}

\bibitem{K-EP}
Lee, D., Hoshi, T., Sogabe, T., Miyatake, Y., Zhang, S.-L.,  
\lq Solution of the k-th eigenvalue problem in large-scale electronic structure calculations', 
J. Comp. Phys. 371 (2018) 618-632.

\bibitem{ELSES-2014}
Hoshi, T., Yamazaki, K., Akiyama, Y., 
\lq Novel linear algebraic theory and one-hundred-million-atom electronic structure calculation on the K computer', 
JPS Conf. Proc. 1 (2014)  016004/1-4.

\bibitem{HOSHI-2009-AUNW}
Hoshi, T., and Fujiwara, T., 
\lq Domain boundary formation in helical multishell gold nanowires', 
J. Phys.: Condens. Matter 21 (2009), 272201/1-7. 

\bibitem{MATRIX-MARKET}
Matrix Market: \url{http://math.nist.gov/MatrixMarket/index.html}





\bibitem{IMACHI-JIT2016}
Imachi, H., Hoshi, T., 
\lq Hybrid numerical solvers for massively parallel eigenvalue computation 
and their benchmark with electronic structure calculations', 
J. Inf. Process. 24 (2016) 164-172.

\bibitem{EIGENKERNEL-URL}
EigenKernel: \verb|https://github.com/eigenkernel/|

\bibitem{EIGENKERNEL-2018}
Tanaka, K., Imachi, H., Fukumoto, T.,  Fukaya, T., Yamamoto, Y., Hoshi, T., 
\lq EigenKernel - A middleware for parallel generalized eigenvalue solvers to attain high scalability and usability',
Preprint: \url{http://arxiv.org/abs/1806.00741}

\bibitem{HOSHI2016SC16}
Hoshi, T., Imachi, H., Kumahata, K., Terai, M., Miyamoto, K., Minami, K., and Shoji F.:
Extremely scalable algorithm for 10$^8$-atom quantum material simulation on the full system of the K computer, 
Proceeding of 7th Workshop on Latest Advances in Scalable Algorithms for Large-Scale Systems (ScalA), 
held in conjunction with SC16:  The International Conference for High Performance Computing, 
Networking, Storage and Analysis Salt Lake City, Utah November, 13-18 (2016) 33-40.


\bibitem{OFET-TEXTBOOK}
Bao, Z., Locklin, J., 
\lq Organic Field-Effect Transistors', CRC Press (2007)


\bibitem{MATSUI-ESR-2010}
Matsui, H. , Mishchenko, A. S., and Hasegawa, T.,
Distribution of localized states from fine analysis of 
electron spin resonance spectra in organic transistors,
Phys. Rev. Lett. 104,  (2010) 056602/1-4.


\bibitem{MATSUI-ESR-2013}
Matsui, H., Mishchenko A. S. and Hasegawa, T. ,
Origin of shallow traps in organic transistors: 
dipole disorder at the semiconductor/insulator interface, 
The 68-th Annual Meeting of the Physical Society of Japan, 
27pXP-4, Hiroshima University, 26-29., Mar. (2013)


\bibitem{GROMACS-URL}
GROMACS: \url{http://www.gromacs.org/}

\bibitem{GROMACS}
Berendsen, H. J. C. . van der Spoel, D., van Drunen, R.,
GROMACS: A message-passing parallel molecular dynamics implementation, 
Comp. Phys. Comm. 91 (1995) 43-56. 


\bibitem{GAFF}
Wang, J., Wolf, R. M., Caldwell, J. W.,
Kollmann, P. A., Case, D. A.,  
Development and Testing of a General Amber Force Field,
Comput. Chem. 25 (2004) 1157-1174.
 

\bibitem{TERAO-2013}
 Terao, J., Wadahama, A., Matono, A., Tada, T., Watanabe, S., Seki, S., Fujihara, T., Tsuji, Y., 
 \lq Design principle for increasing charge mobility of $\pi$-conjugated polymers using regularly localized molecular orbitals',
 Nat. Commun. 4 (2013) 1691.

\bibitem{PCA-REPORT}
Hoshi, T., Imachi, H., Oohira, K.,  Abe, Y., Hukushima, K., 
\lq Principal component analysis with electronic wavefunctions for exploration of organic polymer device materials', 
International Meeting on High-Dimensional Data-Driven Science (HD$^3$-2017), Kyoto, Japan, 10-13, Sep. 2017. 

\bibitem{Z-PARES}
Sakurai, T., Sugiura, H.,  
A projection method for generalized eigenvalue problems using numerical integration, J. Comput. Appl. Math. 159(1) (2003) 119-128.

\bibitem{Z-PARES-URL}
z-PARES: \url{http://zpares.cs.tsukuba.ac.jp/}

\bibitem{FEAST}
Polizzi, E., Density-matrix-based algorithm for solving eigenvalue problems, Phys. Rev. B 79 (2009) 115112/1-6.

\bibitem{FEAST-URL}
FEAST: \url{http://www.feast-solver.org/}

\bibitem{FILTERING}
Li, R., Xi, Y., Vecharynski, E., Yang, C., Saad, Y.,  
A thick-restart Lanczos algorithm with polynomial filtering for Hermitian eigenvalue problems, 
SIAM J. Sci. Comput. 38(4) (2016) A2512-A2534.

\bibitem{K-EP-URL}
k-ep: \url{https://github.com/lee-djl/k-ep}



\end{thebibliography}
\end{document}